\documentclass{aa}
\usepackage{graphicx}
\usepackage{amssymb}
\usepackage{epsfig}
\usepackage{txfonts}

\def\beq{\begin{eqnarray}}
\def\eeq{\end{eqnarray}}
\def\be{\begin{equation}}
\def\ee{\end{equation}}
\def\bc{\begin{center}}
\def\ec{\end{center}}

\def\msun{{\rm M_\odot}}

\def\Te{T_{\rm e}}
\def\d{{\rm d}}
\def\rg{r_g}

\def\Dop{\delta}

\def\vbt{\mbox{\boldmath $\beta$}}
\def\vomega{\mbox{\boldmath $\omega$}}

\begin{document}
   \title{Light curves and polarization of \\
accretion- and nuclear-powered millisecond pulsars}

   \author{K. Viironen
          \inst{}
          \and
          J. Poutanen\inst{}}

   \offprints{J. Poutanen}

   \institute{Astronomy Division, Department of Physical Sciences,
              P.O. Box 3000, FIN-90014 University of Oulu, Finland\\
              \email{kerttu.viironen@oulu.fi, juri.poutanen@oulu.fi}}

\titlerunning{Light curves and polarization from millisecond pulsars}
\authorrunning{K. Viironen \and J. Poutanen }

\date{\today}

\abstract{We study theoretical X-ray light curves and polarization
properties of accretion-powered millisecond pulsars.
We assume that the radiation is produced in  two antipodal
spots at the neutron star surface which are associated with
the  magnetic poles.
We compute the angle-dependent intensity and polarization
produced in an electron-scattering dominated plane-parallel accretion
shock in the frame of the shock.
The observed flux, polarization degree and polarization angle
are calculated accounting for special and general relativistic effects.
The calculations also extended to the case of nuclear-powered
millisecond pulsars -- X-ray bursts. In this case, we consider
one spot and the radiation is assumed to be produced
in the atmosphere of the infinite Thomson optical depth.
The light curves and polarization  profiles  show a large
diversity depending on the model parameters.
Presented results can be used as a first step to understand the
observed pulse profiles of accretion- and nuclear-powered millisecond
pulsars. Future observations of the X-ray polarization will provide
a valuable tool to test the geometry of the emission region
and its physical characteristics.
}

   \maketitle
%

\section{Introduction}

Since the discovery of pulsars in the 1960s, determination of the
equation of state of neutron stars  based on measuring their
mass-radius relation was one of the main research goals.
Radio pulsars have been used extensively for the mass determination
(Thorsett \& Chakrabarty \cite{tc99}), but hardly  can help
to get any constraints on the neutron star radius since the radio
emission is probably produced far from the surface.
Observations of the isolated neutron stars emitting thermal
radiation from their surface on the other hand is much more
suited for the radius determination using the Stefan-Boltzmann law.
Some of the radii determined this way from the X-ray emission
(e.g. Drake et al. \cite{dr02}; Thoma et al. \cite{thoma04})
are much smaller than any of the equation of state allow, pointing
either to the existence of the more compact strange stars or
(more likely) to a surface temperature gradient.

Statistical studies of the pulse profiles from standard X-ray pulsars
also have been used to put constraints on the geometry of the
emission region as well as the star compactness
(Bulik et al. \cite{bul03}). The problem here is that
the accretion rate is close to the (local) Eddington,
the magnetic field strongly affects the radiative transport
resulting in rather complicated pulse profiles at low energies,
and therefore it is difficult to predict from the first
principles the angular distribution of the radiation escaping
the emission region.

Quasi-periodic oscillations (QPO) observed in a number of low-mass
X-ray binaries (van der Klis \cite{vdk00}), if produced in the
inner part of the accretion disk, could be used to put weak
constraints on the neutron star mass as well as its radius (see
e.g. Miller et al. \cite{mlp98}). However, this method is
indirect. Discovery of millisecond coherent pulsations during
X-ray bursts in 13 low-mass X-ray binaries (so called
nuclear-powered millisecond pulsars, see reviews by Strohmayer
\cite{stro01}; Strohmayer \& Bildsten \cite{sb03}) and in the
persistent emission of five  sources (accretion-powered
millisecond pulsars, see review by Wijnands \cite{wij04}) opens a
completely new range of possibilities. The emission in these cases
is produced at the surface of a rapidly spinning neutron star and
the observed pulse profiles and oscillation amplitudes are
strongly affected by special as well as general relativistic
effects  (Braje et al. \cite{bra00}; Weinberg et al. \cite{wei01};
Poutanen \& Gierli\'nski \cite{pg03}, hereafter PG03). Weak
magnetic field in these sources does not affect the radiative
transport and thus a much more reliable prediction for the
radiation pattern from the surface can be obtained. This tightens
the possible range of other parameters. However, similar light
curves still can be produced with a very different set of
parameters. For example, exchanging inclination with the magnetic
inclination (angle between magnetic and rotational axis) has
almost no effect on the light curve.

A possible way to distinguish between the models is to observe the behaviour
of the polarization degree and polarization angle with phase.
Polarization has proven to be a valuable tool in determining the geometry
of the emission region in radio pulsars (e.g. Blaskiewicz et al. \cite{bla91}).
Polarimetric observations in the X-ray range could also be
possible in the near future  (see Costa et al. \cite{costa}; Bellazzini et al.
\cite{bellazzini}; Marshall et al. \cite{marsh03}).
In this paper, we perform theoretical calculations of the light curves
and polarization behaviour in accretion- and nuclear-powered millisecond
pulsars accounting for relativistic effects. The presented models
could be used to determine the physical parameters of the  X-ray
millisecond pulsars.

\section{Method}

We assume that the emission originates in one or two antipodal
(in the case of accreting ms pulsars) spots at the neutron star
surface. The hard X-rays are produced
by thermal Comptonization in a plane-parallel slab (accretion shock)
of optical depth of order unity (PG03).
The emission during the X-ray bursts is assumed to be produced in
a semi-infinite electron scattering dominated atmosphere.
The radiation pattern is assumed to have azimuthal symmetry in the
comoving frame of a spot.
We compute the Stokes parameters in this frame
are transform them to the observer frame. First, we make the
Lorentz transformation to the non-rotating frame accounting for
Doppler boosting and relativistic aberration and then
follow photon trajectories to the observer at infinity
in Schwarzschild space-time.
Deviations from the  Schwarzschild  metric due to the stellar rotation
have a small effect (Braje et al. \cite{bra00}) and are neglected here.
For pulsar rotational frequencies of $\nu\gtrsim 400$ Hz,
we  also need to account for time delays which in the extreme cases
can reach about 5--10 per cent of the pulsar period.

Lorentz transformation and gravitational light bending do not
change the polarization degree and the observed value can be
found if one knows the polar angle at which a photon is emitted in the
spot comoving frame. For a slowly rotating star, the polarization vector
lies in the plane
formed by the spot normal (spot radius-vector) and the line of sight as in
the rotating vector model of Radhakrishnan \& Cooke (\cite{rc69}).
This vector rotates if a star is rapidly spinning
(Ferguson \cite{fer73}, \cite{fer76}).

In the following calculations
we assume two identical antipodal point-like spots,
of which both, or only one is visible depending on the parameters.
We follow the method described in detail in PG03
for  the flux computations and extend it to the description of
the  polarization degree and polarization angle.

\begin{figure}
\begin{center}
\centerline{\epsfig{file=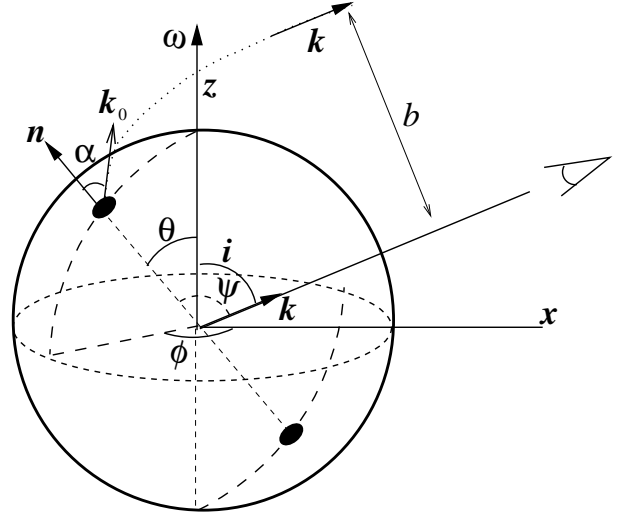,width=8cm}}
\caption{Geometry of the problem.}
\label{fig:geom}
\end{center}
\end{figure}

\subsection{General definitions}

Let $S'$ be a surface element (spot) at colatitude   $\theta$   (see
Fig.~\ref{fig:geom} for the geometry).
(Quantities with primes correspond to the frame comoving with the spot.)
The second spot is situated
at the opposite side of the star.  Let  $i$ be the inclination of the
spin axis and the unit vector in the direction to the observer
$\vec{k}=(\sin i, 0, \cos i)$ lies in the $(x,z)$ plane.
The coordinates of the spots vary periodically and the unit vector of the
spot normal is $\vec{n}=\pm(\sin\theta \cos\phi, \sin \theta \sin\phi,
\cos\theta)$,
where the upper sign corresponds to the spot closer to the observer
(primary) and the lower sign to the antipodal spot (secondary).
Thus the angle between the direction to the spot and the line of sight
is given by
\be \label{eq:psi}
\cos\psi \equiv \vec{k}\cdot\vec{n}
=\pm(\cos i\ \cos\theta+\sin i\ \sin \theta\ \cos\phi),
\ee
where the pulsar rotational phase $\phi$ is zero when the primary spot is
closest to the observer.

In Schwarzschild metric the photon orbits are planar and the
original direction of the photon $\vec{k}_0$ near the stellar surface
is then (PG03):
\be\label{eq:k0}
\vec{k}_0=[ \sin\alpha\ \vec{k} +\sin(\psi-\alpha)\ \vec{n}]/\sin\psi,
\ee
where $\alpha$ is the angle between $\vec{k}_0$ and $\vec{n}$,
i.e. $\cos\alpha=\vec{k}_0\cdot\vec{n}$.

The intensity of radiation and its polarization at a given pulsar phase $\phi$
depend on the angle  $\alpha'$   between an emitted  photon and the local
normal to the stellar surface in the spot comoving frame
\be\label{eq:aberr}
\cos\alpha'=\Dop\ \cos\alpha ,
\ee
where $\Dop$ is the Doppler factor. It can be expressed as
\be \label{eq:dop}
\Dop=1/[\gamma(1-\beta\cos\xi)] ,
\ee
where $\gamma =(1-\beta ^2)^{-1/2}$ and $\beta$ is the spot velocity in units
of speed of light as measured in the non-rotating frame:
\be\label{eq:beta}
\beta=\frac{2\pi R}{c} \frac{\nu}{\sqrt{1-\rg/R}} \sin\theta .
\ee
Here the observed pulsar frequency $\nu$ is corrected for the redshift
$\sqrt{1-\rg/R}$, where $\rg \equiv 2GM/c^2$ is the Schwarzschild radius
of the star of mass $M$ and radius $R$.
The angle $\xi$ between the spot velocity $\vbt/\beta=(\mp\sin\phi ,
\pm\cos\phi ,0)$
and $\vec{k}_0$ can be obtained from Eq.~(\ref{eq:k0}):
\be \label{eq:cosxi}
\cos\xi\equiv \frac{\vbt}{\beta} \cdot \vec{k}_0
=\frac{\sin\alpha}{\sin\psi} \frac{\vbt}{\beta} \cdot \vec{k}=
\mp \frac{\sin\alpha}{\sin\psi}\sin i\ \sin\phi\  .
\ee

\subsection{Light bending and time delay}

Due to the gravitational bending of light the angle at which the radiation is
observed is different from the actual emission angle (see
Fig.~\ref{fig:geom}).
The relation between angles $\alpha$ and $\psi$ can be obtained by
computing an elliptical integral (Pechenick et al. \cite{pfc83}):
\be \label{eq:bend}
\psi=\int_R^{\infty} \frac{\d r}{r^2} \left[ \frac{1}{b^2} -
\frac{1}{r^2}\left( 1- \frac{\rg}{r}\right)\right]^{-1/2} ,
\ee
where the impact parameter
\be \label{eq:impact}
b=\frac{R}{\sqrt{1-\rg/R}} \sin\alpha .
\ee
For many purposes one can use a simpler expression derived
by Beloborodov (\cite{belo02}, hereafter B02):
\be \label{my1}
\cos\alpha \approx \cos \psi \left (1-\frac{\rg}{R}\right )+\frac{\rg}{R}\ ,
\ee
which gives an accuracy of the order of a few percent for $R \gtrsim 2.5\rg$.
The spot is visible when $\cos\alpha > 0$. In Beloborodov's approximation
this corresponds to $\cos\psi > - \rg/(R-\rg)$.

The difference in light travel time around  the star
becomes larger for more compact stars.  A photon of impact parameter $b$
is lagging one with the impact parameter $0$ by
(Pechenick et al. \cite{pfc83}):
\be \label{eq:delay}
c\Delta t(b)=  \int_R^{\infty} \frac{\d r}{1- \rg/r}
\left\{ \left[ 1-  \frac{b^2}{r^2}  \left( 1- \frac{\rg}{r} \right)
\right] ^{-1/2}  -1 \right\} .
\ee
For rough estimations one can use an approximation
$\Delta t(b)\approx R (1-\cos\psi) /c$.
If we compute the phase delays relative to photons
emitted at $\phi=0$, then the observed phase is approximately
\be
\phi_{\rm obs} \approx \phi + \Delta \phi_0 (1-\cos\phi),
\ee
where
\be
\Delta \phi_0=  \cos i\ \cos\theta\ 2\pi \nu R/c \ .
\ee
For small delays, this expression can be reversed to
\beq \label{eq:lightdel}
\cos \phi & \approx & \cos \phi_{\rm obs} + \Delta \phi_0 \ \sin\phi_{\rm obs}
(1-\cos\phi_{\rm obs}) , \\
\sin \phi & \approx & \sin \phi_{\rm obs} + \Delta \phi_0 \ \cos\phi_{\rm obs}
(1-\cos\phi_{\rm obs}) . \nonumber
\eeq

\subsection{Observed intensity and flux}

The combined effect of the gravitational redshift and Doppler effect
results in the following  relation between the monochromatic
observed and local intensities
(see e.g. Misner et al. \cite{mtw73}; Rybicki \& Lightman \cite{rl79}):
\be
I_{E} = \left (\frac{E}{E'}\right )^3 I'_{E '} (\alpha')
\ee
where $E/E'=\Dop \sqrt{1-r_g/R}$.
Here $I'_{E'}(\alpha')$ is the intensity computed in the frame comoving with
the spot.
For the bolometric intensity, one gets
\be
I= \left (\Dop \sqrt{1-r_g/R} \right )^4 I'(\alpha') .
\ee

If the radiation spectrum can be represented by
a power-law $I'_{E'}(\alpha') \propto E '^{-(\Gamma-1)}$
(as in accretion-powered ms pulsars)
with the photon spectral index $\Gamma$ which does not depend on the
angle $\alpha'$, then
\be
I'_{E'}(\alpha') = I'_{E}(\alpha')
\left( \Dop \sqrt{1-\rg/R} \right)^{\Gamma-1} .
\ee
This approximation is equivalent to the assumption of a weak
energy dependence of the angular distribution. In the case of thermal
Comptonization, it can be used if the maximum Doppler shift
$\Delta \delta \sim 2\pi \nu R/c \ \sin i\ \sin\theta$
 is smaller than the typical (relative) energy separation between photons of
subsequent scattering orders $\Delta E/E \sim 4k\Te/m_{\rm e} c^2$,
where $T_{\rm e}$ is the electron temperature. This condition is equivalent to
$ (\nu/600\ \mbox{Hz}) \sin i\ \sin\theta < k\Te/16\ \mbox{keV}$, which
is always satisfied in accreting pulsars.
Then, the observed photons at a given energy $E$ belong mostly to a
certain scattering order ($n=0,1,2,...$) which can be determined from
\be\label{eq:ene_nscat}
\frac{E}{E_0} = \left( 1+
\frac{4k\Te}{m_{\rm e} c^2} \right)^n ,
\ee
where $E_0$ is the mean energy (at infinity) of the seed photon
black body distribution.

In order to compute the observed flux, we need to know variations
of the solid angle occupied by the spot on the observer's sky (PG03):
\be\label{eq:omega}
\Omega=\frac{S' \cos \alpha'}{D^2}\ \frac{1}{1-\rg/R}\
\frac{\d\cos\alpha}{\d\cos\psi}.
\ee
where $D$ is the source distance. (This transforms to
$\Omega \approx S' \cos\alpha' /D^2$ in Beloborodov's approximation,
since $\d\cos\alpha/\d\cos\psi\approx 1-\rg/R$.)
Hence the monochromatic flux becomes:
\be \label{eq:fluxmono}
F_{E}=I_E \Omega=
(1-\rg/R)^{1/2} \Dop^{4} I'_{E'}(\alpha') \cos\alpha
\frac{\d\cos\alpha}{\d\cos\psi}
 \frac{S'}{D^2} ,
\ee
where we used aberration formula (\ref{eq:aberr}).
The bolometric flux is given by:
\be \label{eq:fluxbolo}
F= (1-\rg/R)\ \Dop^5 \
I'(\alpha')  \cos\alpha \frac{\d\cos\alpha}{\d\cos\psi} \frac{S'}{D^2} .
\ee
For a power-law spectrum, the monochromatic flux is
\be\label{eq:fluxplaw}
F_{E}= (1-\rg/R)^{(\Gamma+2)/2}\ \Dop^{\Gamma+3} I'_E(\alpha')
\cos\alpha \frac{\d\cos\alpha}{\d\cos\psi} \frac{S'}{D^2} .
\ee
In the following calculations we have used Eqs.~(\ref{eq:fluxbolo})
and~(\ref{eq:fluxplaw}) (with $\Gamma = 2$)
for the black body and power-law spectra, respectively.
We normalize the fluxes with
$F_0 = 2(1-r_g/R)^2 \frac{S'}{D^2} \int _0^1 \mu I'(\mu)d\mu $.

\subsection{Polarization degree and polarization angle}

Relativistic effects do not change the polarization degree and thus we
take it equal to the polarization in the spot comoving frame
$P'_{E'}(\alpha')$.
However, when calculating the polarization degree of the total flux
from both spots, the difference in the tilt of the polarization ellipses
(difference caused by the time delay and different relativistic rotations)
need to be considered.

In order to describe polarization, we introduce the main polarization
basis,
\be
\vec{e}_1^{\rm m}=\frac{\hat{\vomega} - \cos i\ \vec{k} }{\sin i}, \quad
\vec{e}_2^{\rm m}=\frac{\vec{k} \times \hat{\vomega}}{\sin i} ,
\ee
where $\hat{\vomega}$ denotes the unit vector
along the stellar rotational axis.
In the absence of relativistic rotation of the polarization plane (for a
slowly rotating star), the polarization vector lies in the plane formed by
the local normal $\vec{n}$ and the direction to the observer $\vec{k}$.
The corresponding polarization basis is
\be
\vec{e}_1=\frac{\vec{n} - \cos \psi\ \vec{k} }{\sin \psi}, \quad
\vec{e}_2=\frac{\vec{k} \times \vec{n} }{\sin \psi} .
\ee
The polarization angle (PA) $\chi_0$ measured from the projection of the
spin axis on the plane of the sky in the counter-clockwise direction
is given by:
\beq\label{eq:coschi}
\cos \chi_0 & = & \vec{e}_1^{\rm m} \cdot \vec{e}_1 = \vec{e}_2^{\rm m} \cdot \vec{e}_2
 =  \frac{\sin i\ \cos \theta - \cos i\  \sin \theta\  \cos \phi }
{\sin \psi}, \nonumber \\
\sin \chi_0 & = & \vec{e}_2^{\rm m} \cdot \vec{e}_1 = - \vec{e}_1^{\rm m} \cdot \vec{e}_2
= - \frac{\sin \theta\ \sin \phi} {\sin \psi} .
\eeq
We thus get
\be \label{eq:papr}
\tan\chi_0=-\frac{\sin \theta\ \sin \phi}
{\sin i\ \cos \theta  - \cos i\ \sin \theta\  \cos \phi }.
\ee
This expression neglects the rotation of the polarization plane
due to relativistic motion (e.g. Ferguson \cite{fer73}, \cite{fer76}).
To correct for that we introduce the
polarization basis related to the photon direction close
to the neutron star surface $\vec{k}_0$:
\be
\vec{e}_1^0=\frac{\vec{n} - \cos \alpha\ \vec{k}_0 }{\sin \alpha}, \quad
\vec{e}_2^0=\frac{\vec{k}_0 \times \vec{n} }{\sin \alpha} =  \vec{e}_2.
\ee
The equality $\vec{e}_2^0=\vec{e}_2$ is related to the fact that photon
trajectories are planar.
If the polarization vector in the spot rest frame lies in the meridional
plane, then it will be transformed to (Nagirner \& Poutanen \cite{np93})
\be
\vec{e}'_1=\frac{\vec{n} +\Dop \gamma\ \cos \alpha\ (\vbt- \vec{k}_0 ) }
{\sqrt{1-\Dop^2 \cos^2\alpha}}
\ee
in the non-rotating frame.
Thus, the rotation of the polarization vector due to relativistic
motion is given by
\be
\cos \chi_{\rm c} = \vec{e}'_1 \cdot \vec{e}_1^0 \ , \quad
\sin \chi_{\rm c} = \vec{e}'_1 \cdot \vec{e}_2^0 \ ,
\ee
which is reduced to (Poutanen, in preparation)
\be\label{eq:pac}
\tan \chi_{\rm c}=\beta \cos\alpha \frac{\cos i\ \sin \theta -
\sin i\ \cos \theta\ \cos \phi}
{\sin \alpha\ \sin\psi +\beta \sin i\ \sin\phi} .
\ee
This correction angle depends linearly on the spot velocity $\beta$ and,
therefore, is negligible for spin frequencies $\nu \lesssim  300$~Hz.
Since the angle $\chi_{\rm c}$ is constant for parallel transport along
the photon trajectory, the total polarization angle for each spot is
\be \label{eq:pa}
\chi =   \chi_0 + \chi_{\rm c} .
\ee
When we plot polarization angles for individual spots,
we rotate it by $\pi/2$ (in any direction) if
the polarization $P=Q/I$ is  negative (i.e. dominant electric vector
oscillations are perpendicular to the meridional plane
defined by the spot normal and the direction of
photon propagation).

For each spot calculations provide us with the Stokes vector
$(F_I,F_Q,0)^{\rm T} $
(we do not consider sources of circular polarization), where
flux $F_I$ is computed using  Eqs.~(\ref{eq:fluxbolo}) or (\ref{eq:fluxplaw}),
$F_Q=P F_I$,  and the polarization degree $P$ at angle $\alpha'$
is computed for a given optical depth of the slab (see below
Eqs.~(\ref{eq:polslab}) and  (\ref{eq:polburst})).
We rotate this vector to the main basis
\be
\left( \begin{array}{ccc}  1 & 0 & 0 \\ 0 & \cos2\chi & -\sin2\chi \\
0 & \sin2\chi & \cos2\chi  \end{array}  \right)
\left(\begin{array}{c}   F_I \\ F_Q \\ 0  \end{array} \right)=
\left(\begin{array}{c}   F_I  \\ F_Q\cos2\chi \\ F_Q\sin2\chi  \end{array}
\right) .
\ee
Denoting the corresponding quantities for the primary and secondary spots
by indicies ${\rm p}$ and ${\rm s}$, respectively, we get for the
total observed Stokes vector
\be
\left( \begin{array}{c}
F_I^{\rm p} + F_I^{\rm s}  \\
F_Q^{\rm p} \cos2\chi_{\rm p} + F_Q^{\rm s} \cos2\chi_{\rm s}  \\
F_Q^{\rm p} \sin2\chi_{\rm p} + F_Q^{\rm s} \sin2\chi_{\rm s}
\end{array} \right) .
\ee
The degree of polarization is then obviously
\be\label{eq:poltot}
P = \frac{\sqrt{(F_Q^{\rm p})^2+(F_Q^{\rm s})^2 + 2F_Q^{\rm p}F_Q^{\rm s}
\cos (2\chi_{\rm p}- 2\chi_{\rm s})}}{F_I^{\rm p}+F_I^{\rm s}},
\ee
and the total polarization angle is given by
\be \label{eq:patot}
\tan 2\chi_{\rm tot} = \frac{F_Q^{\rm p} \sin 2\chi_{\rm p} +
F_Q^{\rm s}  \sin 2\chi_{\rm s}}
{F_Q^{\rm p} \cos 2\chi _{\rm p} + F_Q^{\rm s} \cos 2\chi_{\rm s}} .
\ee


\section{Local intensity and polarization}
\label{sec:thomson}

\begin{figure*}
\centering{\epsfig{figure=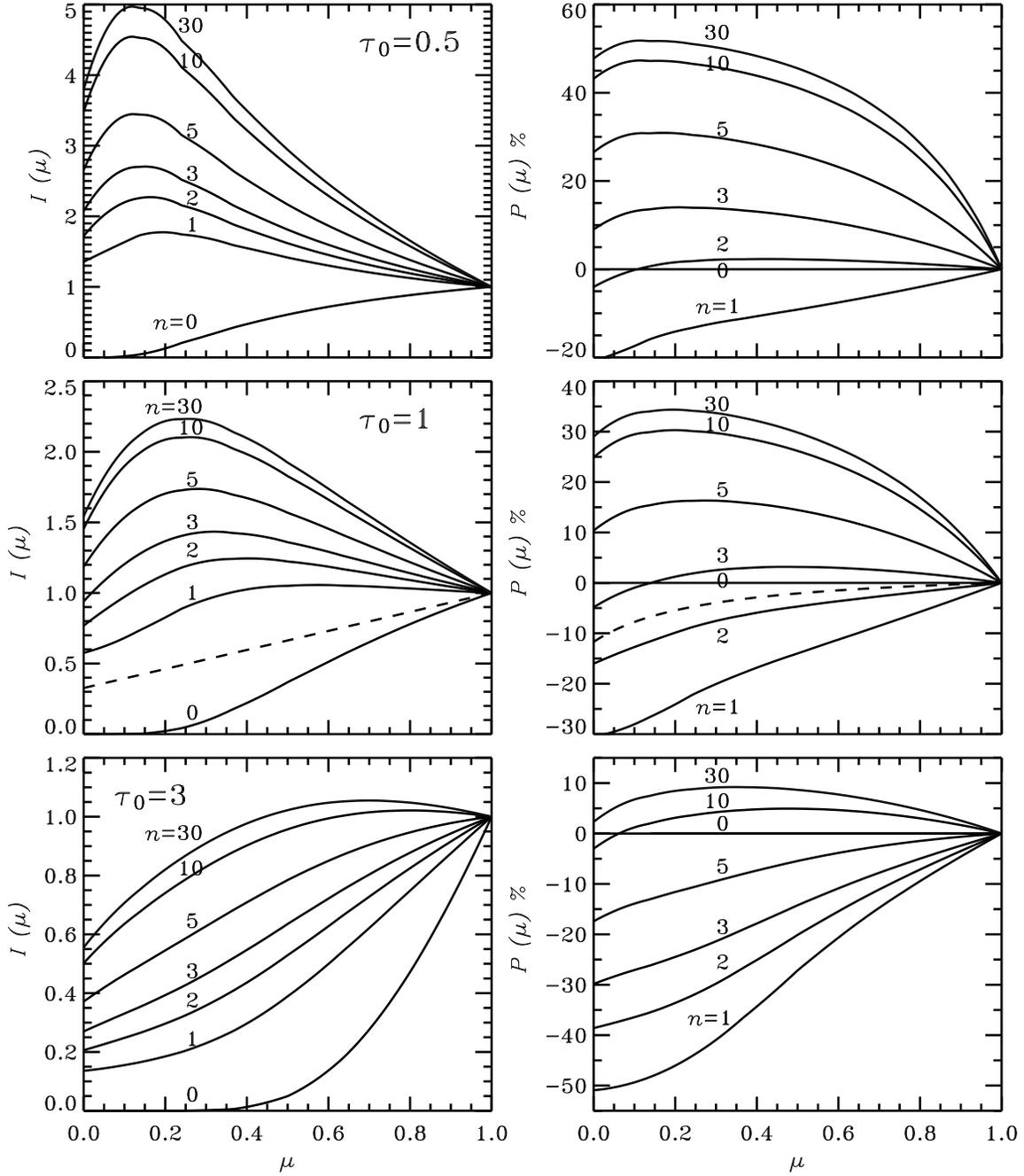,width=0.85\linewidth}}
\caption{Intensity (normalized to unity at $\mu=1$)
and polarization of the radiation
escaping from a slab of Thomson optical depth  $\tau_0=0.5, 1, 3$
for different scattering orders $n$.
The unpolarized seed photons with intensity $I(\mu)=1$ are injected
from the bottom of the slab.
The dashed curves show the intensity and polarization
corresponding to the classical results of Chandrasekhar-Sobolev for the
semi-infinite atmosphere $\tau_0=\infty$.
}
\label{fig:intpol}
\end{figure*}

\subsection{Radiative transfer in the accretion shock}

We assume that the accretion shock can be
represented as a plane-parallel slab which horizontal size
is much larger than its vertical height
so that the photons escaping from the sides of the slab can be ignored.
This is a good approximation for moderate luminosities
(Basko \& Sunyaev \cite{bs76}) typical for accretion-powered
millisecond pulsars (Wijnands \cite{wij04}).
We calculate the angle-dependent intensity and the polarization
degree of radiation emerging from the emission region, assuming
Thomson scattering and corresponding polarization.
The calculations are carried out following the procedure described
in Sunyaev \& Titarchuk (\cite{st85}).
Unpolarized seed soft photons are assumed to be injected from
the neutron star surface  (i.e. bottom of the slab).
The boundary conditions are
\be \label{bound1}
I_{l,r}(\tau =0,\mu) = I_{l,r}(\tau =\tau _0, -\mu) = 0,
\quad 0 \leq \mu\equiv \cos\theta \leq 1 ,
\ee
where $\theta$ is the angle between photon direction and the outward
slab normal, indices $l,r$ refer to the polarizations in the
meridional plane and perpendicular to it, respectively.
The Thomson optical depth $\tau$ is measured from the bottom
of the slab.  The intensity of unscattered photons  is given by:
\be
I_{l,r}^0(\tau,\mu) =  I_0 {\rm e}^{-\tau/\mu}.
\ee
The intensity of $n$ times scattered photons can be computed from
\beq\label{intensity2}
I_l^n(\tau,\mu)& = & \int_0^{\tau} \exp \left (-\frac{\tau -\tau '}{\mu}
\right )
\nonumber \\
&\times & \left\{ (1-\mu ^2)A^{n}(\tau ') +\mu ^2\left[B^{n}(\tau ')+
C^{n}(\tau ')\right]\right\} \frac{\d\tau '}{\mu}\,,  \nonumber \\
I_r^n(\tau,\mu) & =& \int_0^{\tau}\exp \left(-\frac{\tau -\tau '}{\mu}\right)
\left[B^{n}(\tau ')+C^{n}(\tau ')\right]\frac{\d\tau '}{\mu}\,,
\eeq
where the quantities $A^{n}(\tau)$, $B^{n}(\tau)$ and $C^{n}(\tau)$ are
as follows
\beq
A^{n}(\tau) &=& \frac{3}{4}\int_0^1(1-\mu ^{\prime 2})\left[
I_l^{n-1}(\tau,\mu ')+I_l^{n-1}(\tau,-\mu ')\right] \d\mu ', \nonumber \\
B^{n}(\tau) & =& \frac{3}{8}\int_0^1\mu ^{\prime 2}\left
[I_l^{n-1}(\tau,\mu ')
+I_r^{n-1}(\tau,-\mu')\right ]\d\mu ', \\
C^{n}(\tau) &= &\frac{3}{8}\int_0^1\left [I_r^{n-1}(\tau,\mu ')+
I_r^{n-1}(\tau,-\mu ')\right ]\d\mu ' .\nonumber
\eeq
The intensities $I_{l,r}^n(\tau,-\mu)$ can be computed in a similar way
by changing the integration boundaries in Eqs. (\ref{intensity2})
from $(0,\tau)$ to $(\tau, \tau _0)$. These equations are similar
to those in Sunyaev \& Titarchuk (\cite{st85}),
but we do not assume symmetrical (relative to the slab central plane)
distribution of seed photons.
Finally, the degree of
polarization of the radiation leaving the slab upper boundary is
\be \label{eq:polslab}
P = Q(\tau _0,\mu)/I(\tau _0,\mu) ,
\ee
where the Stokes parameters are
\beq \label{inteq}
I(\tau _0,\mu)&=& I_l(\tau _0,\mu) + I_r(\tau _0,\mu),  \nonumber \\
Q(\tau _0,\mu)&=& I_l(\tau _0,\mu) - I_r(\tau _0,\mu) .
\eeq
The results of calculations
for different scattering orders are presented in Fig.~\ref{fig:intpol}.

Due to the symmetry, the polarization  vector can be either in
the meridional plane (positive polarization)
or perpendicular to it (negative polarization).
The polarization degree is zero for the  seed radiation ($n=0$),
negative for small scattering orders and for $\tau_0=1$ becomes positive at
$n \gtrsim 3$.
The polarization degree also depends on the electron temperature.
At high electron temperatures,  the electron random motions produce
different aberrations and in the electron rest
frame  photons are scattered  at different  angles.  This
reduces the  polarization degree (Poutanen 1994; Poutanen \& Svensson 1996).
Thus, for example, for $k\Te\sim 100$ keV, polarization is two times
smaller than that given in the Thomson scattering approximation.

Variation of the optical depth also strongly affects the angular distribution
of radiation and the polarization degree. An increase in $\tau_0$
leads to more beaming along the normal to the slab
and to smaller $P$ (Sunyaev \& Titarchuk \cite{st85}).
At larger $\tau_0\sim 3$ and $n\gg1$ (see $n=10$ curve at the
lower panel in Fig.~\ref{fig:intpol})
the polarization changes the sign for some angles reminding the behavior of $P$
for $\tau_0=1$ and $n=3$.

Since in the process of thermal Comptonization the energy of
a photon  increases with the number of scattering, there is one-to-one
correspondence between scattering order and the energy.
For our further calculations we choose optical depth $\tau _0 = 1$
since the X-ray spectra of accreting neutron stars
at low and intermediate luminosities  are often described
by thermal Comptonization in a slab of optical depth of order unity
and electron temperature $k\Te\sim 50$ keV
(Gilfanov et al. \cite{gil98}; Barret et al. \cite{bar00}; PG03).
Polarization degree for larger and smaller $\tau_0$ can be estimated
from Fig.~\ref{fig:intpol}.

\begin{figure}
\begin{center}
\centerline{\epsfig{file=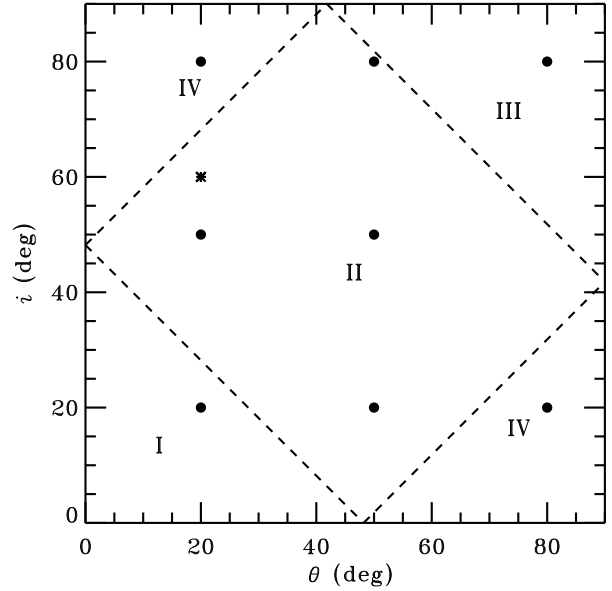,width=0.9\linewidth}}
\caption{Classes of the light curves for two antipodal spots and
$R=2.5\rg$ according to Beloborodov (2002).
Asterisk shows our basic parameter set
and circles  correspond to other considered cases.
}
\label{fig:class}
\end{center}
\end{figure}

\subsection{X-ray bursts}

If during the X-ray bursts, we can observe coherent pulsations,
the atmosphere is pinned down to the neutron star surface, and thus
a plane-parallel approximation is still valid.
In a classical problem of radiative transfer in a plane-parallel
semi-infinite ($\tau _0 \rightarrow \infty$) atmosphere with opacity
dominated by Thomson scattering (Chandrasekhar \& Breen \cite{cha47};
Chandrasekhar \cite{cha60}; Sobolev \cite{sob49};   Sobolev \cite{sob63}),
the intensity of escaping radiation is approximated
by the formula $I(\mu) \sim 1+2.06\mu$.
The degree of polarization is at its maximum, $11.7\,\%$, at $\mu = 0$ and
decreases to $0 \%$ at $\mu = 1$ due to the symmetry.
The polarization degree can be approximated as
\be \label{eq:polburst}
P=-\frac{1-\mu}{1+3.582\mu} 11.71\% .
\ee
The dominant direction of the electric vector oscillations
is perpendicular to the meridional plane.

\section{Results}

\subsection{Pulse profiles for black body spots}
\label{sec:bb}

\begin{figure*}
\centering{\epsfig{figure=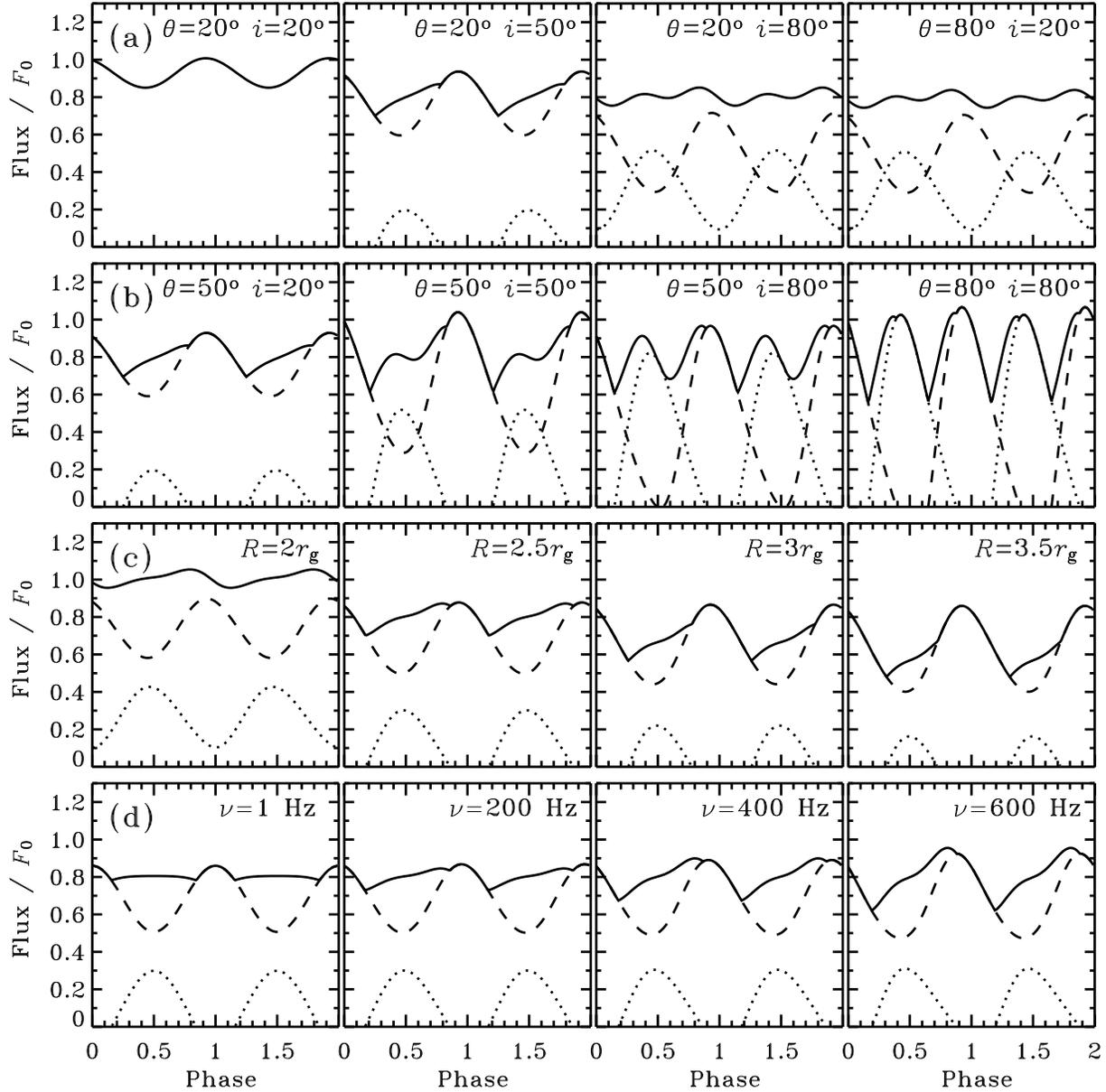,width=0.90\linewidth}}
\caption{Observed flux from two antipodal  black body spots as a function
of phase $\phi_{\rm obs}/(2\pi)$.
The basic set of parameters is  $\theta=20\degr$, $i=60\degr$, $\nu=300$ Hz,
$R=2.5\rg$ and $M=1.4\msun$. {\bf a)}, {\bf  b)} Dependence on the
 inclination angle $i$ and the magnetic angle $\theta$ with other parameters
from the basic set.
Dependence on the neutron star compactness is shown in panels {\bf (c)},
while panels {\bf (d)} show the effect of different spin frequency.
Solid curves correspond to the total flux, while the
dashed curves are  for the primary spot, and
dotted curves -- for the secondary.}
\label{fig:fluxbb}
\end{figure*}

The light curves produced by two small antipodal spots emitting as a
black body can be divided into four classes (B02, see Fig.~\ref{fig:class}):
I --  only one spot is visible, II --  the primary spot is always visible,
the secondary spot is visible sometimes,
III -- both spots are eclipsed sometimes,
IV --  both spot are visible all the time.
The light curve from a single spot in Beloborodov's approximation
can be represented as follows (PG03; Poutanen \& Beloborodov, in preparation):
\beq \label{eq:fluxapp}
F &\propto &\delta^5 [\rg/R+(1-\rg/R)\cos\psi ]  \nonumber \\
&\approx& \left(1-5\beta\sqrt{1-\rg/R}\ \sin i\ \sin\phi\right) (Q+U\cos\phi) ,
\eeq
where
\beq \label{eq:UQ}
U&=&(1-\rg/R)\sin i\ \sin\theta, \nonumber \\
Q&=&\rg/R+(1-\rg/R)\cos i\ \cos\theta .
\eeq
Here we took into account only the first term of the expansion
of the Doppler factor  in velocity $\beta$
and approximated $\sin\alpha/\sin\psi$ by
its asymptotic value for small angles $\sqrt{1-\rg/R}$.
Thus for small $\beta$, the observed flux varies as a sinusoid. At high
rotational frequencies, the oscillations are shifted in phase and a harmonic
appears as a result of Doppler effect.
An additional source of the phase shift and the signal at the harmonic
is the light travel delays (see Eq.~(\ref{eq:lightdel})).
One can notice that the exchange of $i$ and $\theta$ does not affect
the light curve (for small $\beta$) since the flux depends on the combinations
$\sin i\ \sin\theta$ and $\cos i\ \cos\theta$ (see Eqs.~(\ref{eq:beta}),
(\ref{eq:lightdel}), (\ref{eq:fluxapp}),
(\ref{eq:UQ})).

In Fig.~\ref{fig:fluxbb}a,b we plot the light curves for
different couples of $i$, $\theta$ shown in Fig.~\ref{fig:class}.
The phase shift due to the light travel effect
is computed relative to the photons emitted at $\phi=0$ (so that
the observed phase $0$ coincides with the pulsar phase $\phi=0$).
For a given $\theta$, the variability amplitude from a single spot is
increasing with $i$ since polar angle $\alpha$ covers a larger interval.
The semi-amplitude is simply
\be \label{eq:ampl1}
A \equiv (F_{\max}-F_{\min})/(F_{\max}+F_ {\min})\approx  U/Q.
\ee

\begin{figure*}
\begin{center}
\centerline{\epsfig{file=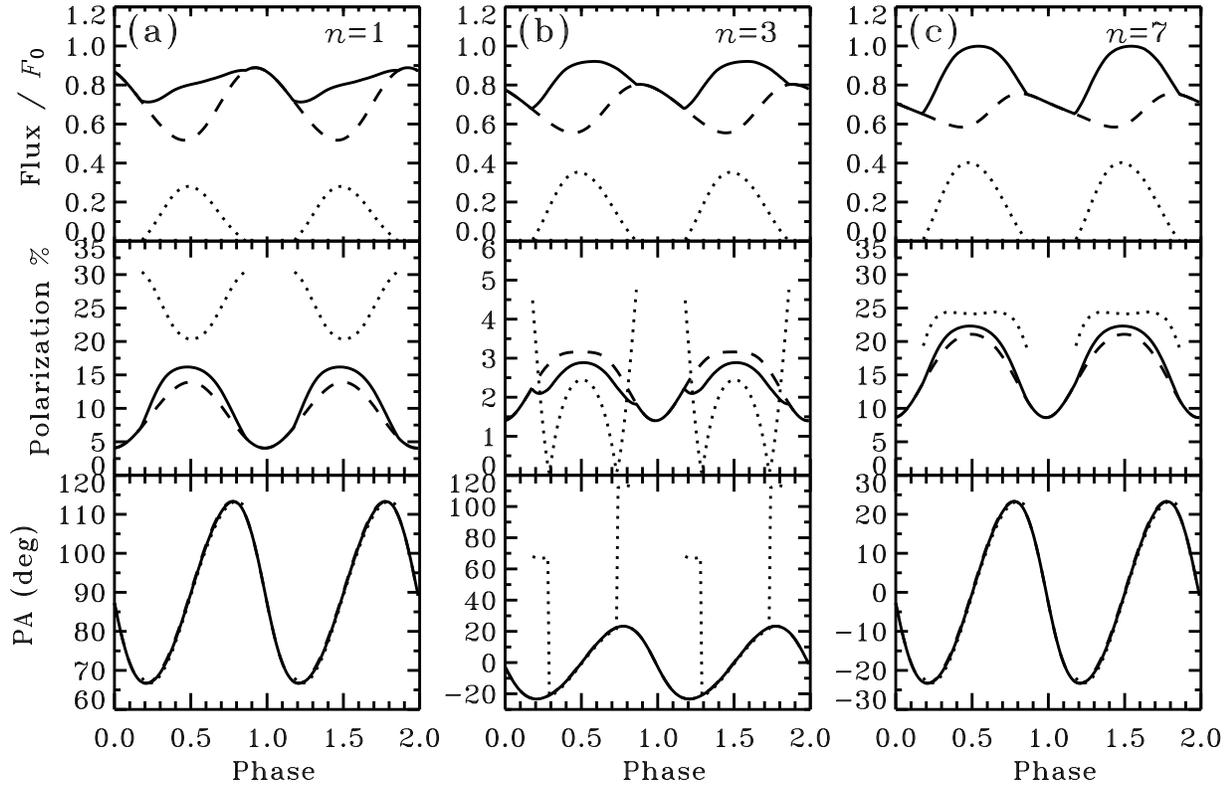,width=0.9\linewidth}}
\caption{Pulse profiles of the flux, polarization degree, and polarization
angle of the scattered radiation for different scattering orders.
The emission region is a slab of Thomson optical depth $\tau _0 = 1$.
The same basic set of parameters and the same designations for the curves
are used as in Fig.~\protect\ref{fig:fluxbb}.
}
\label{fig:nscat}
\end{center}
\end{figure*}

The three panels on the top raw show the effect of varying
inclination only.
At small inclinations, only primary spot is visible (class I)
and deviations from the sine wave are small.
At larger inclination, the second
spot appears at some moments (class II), when the flux from the
first spot is at minimum. The total variability amplitude
then decreases. For a  slowly rotating star, the increase in the
second spot flux would exactly cancel the decrease of the
flux from the first spot making a plateau in the total flux.
This does not happen when
the star rotates at 300 Hz. Rapid rotation introduces an asymmetry,
boosting the radiation when the spot is moving towards the observer.
Since the peak of the  Doppler factor reaches
the maximum quarter of the period earlier than the
projected area, the peak of the light curve shifts to an earlier phase.
At very large inclinations (third panel a),
both spots are visible all the time. The light curve belongs to the
class IV and in the absence of rotation would be flat (B02).
The variability amplitudes at the fundamental frequency and its harmonic,
proportional to $\propto \nu\ \sin 2i\ \sin2\theta$ and to
$\propto \nu\ \sin^2i\ \sin^2\theta$, respectively (Poutanen \&
Beloborodov, in preparation), are small.

The light curves for other values of inclination $i$ and
the magnetic angle $\theta$ are shown in Fig.~\ref{fig:fluxbb}b.
One sees that exchange $i$ and $\theta$ has no effect on the light curve
 as was noticed above on theoretical grounds (see Eqs.~(\ref{eq:fluxapp}),
(\ref{eq:UQ})).
The variability amplitude for individual spots increases with
$\theta$ since it is approximately proportional to $\sin\theta$
as can be seen from Eq.~(\ref{eq:ampl1}).
At very large $i$ and $\theta$ (rightmost panel Fig.~\ref{fig:fluxbb}b),
both spots are sometimes eclipsed.
Two plateaus, that would appear for slowly rotating stars when both
spots are visible, are not present when rotation is fast.
Two pulses corresponding to the two spots are almost identical
producing a strong signal at a harmonic frequency.

Figure~\ref{fig:fluxbb}c demonstrates the dependence of the
light curves on the neutron star compactness $M/R$.
For large compactness, i.e. small radius $R\lesssim 2\rg$,
both spots are always visible (class IV) for most of the
parameters $i$, $\theta$.
The variability amplitude is a linear function of the spot velocity.
For larger radii,  the second spot  becomes eclipsed
(for the set of parameters chosen),
and the variability amplitudes grows. This can be easily understood
from Eq.~(\ref{eq:ampl1}).

The dependence on the rotational frequency is shown in Fig.~\ref{fig:fluxbb}d.
For very slow rotation speed the individual profiles are close to
pure sinusoids and the total profile is symmetric with
flat parts at the phases were both spots are visible (B02).
The larger is the spin frequency, the more the individual profiles
become skewed and the more distorted is the total profile. The individual
pulse amplitudes also slightly increase with increasing frequency due to
variations in the Doppler factor.

\subsection{Pulse profiles and polarization of the Comptonized emission}

\begin{figure*}
\begin{center}
\centerline{\epsfig{file=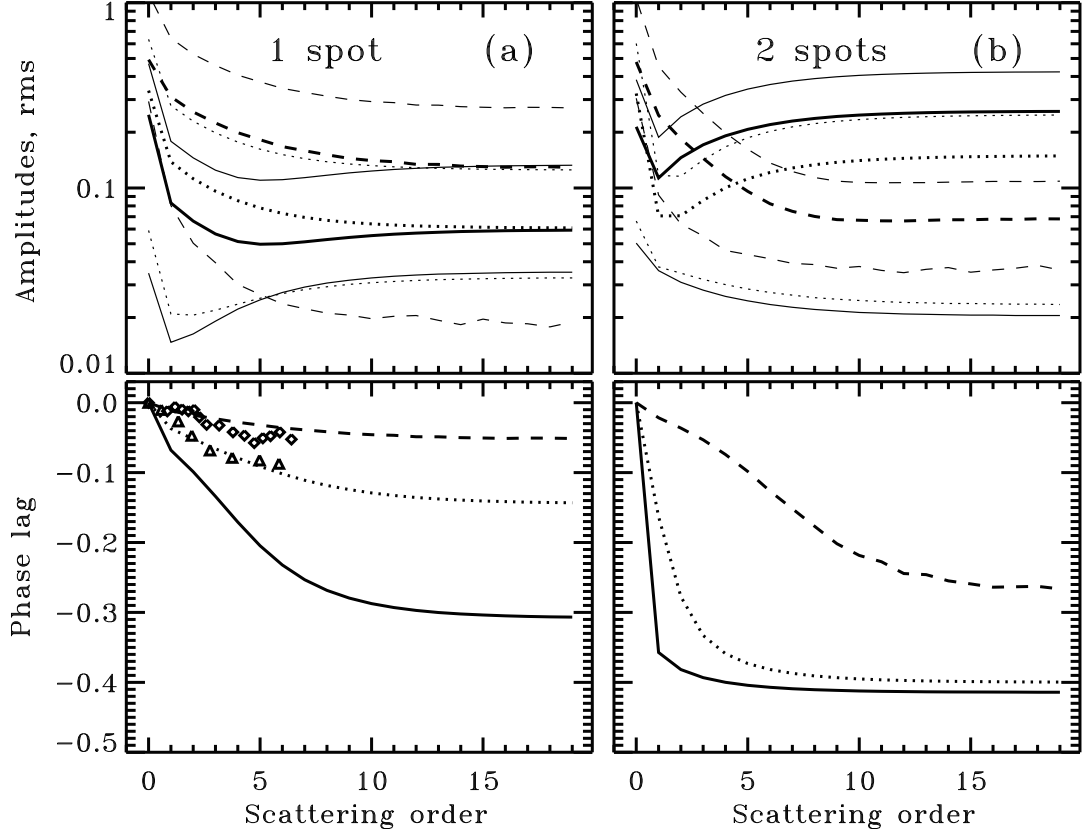,width=0.8\linewidth}}
\caption{
The amplitudes of the fundamental $a_1(n)$ and the first
harmonic $a_2(n)$  (thin curves, fundamental has a larger amplitude)
and the total rms (thick curves) as a function of the scattering order
are shown in the top panels. Solid, dotted, and dashed curves
correspond to the slab optical depth $\tau_0=0.5, 1, 3$, respectively.
The model phase lags at the fundamental frequency relative to $n=0$,
$\phi_1(n)-\phi_1(0)$ (thick curves), and those at the first
harmonic, $\phi_2(n)-\phi_2(0)$ (thin curves), are shown in the lower panels.
Left panels {\bf (a)} correspond to one spot and right panels {\bf (b)}
to two antipodal spots.
The parameters are: $i=60\degr$, $\theta=20\degr$,
$R=2.5\rg$, $M=1.4\msun$. The pulsar frequency was chosen $\nu=401$ Hz
which corresponds to that of SAX~J1808.4$-$3658.
The observed phase lags from that source are shown in lower panel (a) by
triangles.
We transformed the photon energy to the scattering order using
Eq.~(\ref{eq:ene_nscat}) with $k\Te=50$ keV and $E_0=2.8$ keV.
Diamonds represent the phase lags observed in the 2.3 ms (435 Hz)
accreting pulsar XTE~J1751$-$305 (Gierli\'nski \& Poutanen, in preparation).
}
\label{fig:phaselag}
\end{center}
\end{figure*}

The spectra of accreting millisecond pulsars can be represented
as a sum of a black body-like emission
and a Comptonized tail (Gilfanov et al. \cite{gil98};
Gierli\'nski et al. \cite{gier02}; PG03; Markwardt et al. \cite{mark02};
Galloway et al. \cite{gal02}).
The black body photons play also a role of seed photons for
Comptonization.  Since below a few keV contribution of the black body
is large the expected polarization degree is small (unless this emission
is a Wien peak from electron-scattering dominating atmosphere which can
be polarized). Scattering in the hot electron slab modifies significantly
the angular distribution of radiation and
produces polarization signal as described in \S~\ref{sec:thomson}
(see also Poutanen \& Svensson \cite{ps96}). The polarization degree is
a strong function of  the scattering order for small $n$
(see Fig.~\ref{fig:intpol}) and therefore the energy (close to the peak of
the black body). After 7--10 scatterings polarization saturates.
For most of the simulations below we use $n=7$.
This would correspond to about 10 keV for a typical electron
temperature of 50 keV and the seed photon temperature of 0.6 keV.
At higher $\Te$, the corresponding energy is higher, while
polarization at 10 keV is then smaller. Increase of the electron
temperature generally have a depolarizing effect (Poutanen \cite{p94}).
We choose the slab optical depth $\tau_0=1$ which is consistent with the
spectra of SAX J1808.4$-$3658 (PG03).
The predicted polarization degree anti-correlates with the value of $\tau_0$
(a smaller $\tau_0$ would increase it and vice versa).

Since the polarized flux corresponds to a power-law like Comptonized
spectrum, we use Eq.~(\ref{eq:fluxplaw}) for the calculations.
The polarization degree for individual spots
is determined by the polar angle $\alpha'$ in
the spot comoving frame and we plot its absolute value.
The total polarization degree is given by Eq.~(\ref{eq:poltot}).
The polarization angle is given by Eqs.~(\ref{eq:papr})--(\ref{eq:pa}).
If the polarization degree
$P(\alpha')$ is negative (in our convention this corresponds
to the electric vector oscillations perpendicular to the
meridional plane formed by a photon momentum and a local normal
to the slab), we rotate the PA for individual spots by $90\degr$.
The total polarization angle is defined by Eq.~(\ref{eq:patot}).

\begin{figure*}
\begin{center}
\centerline{\epsfig{file=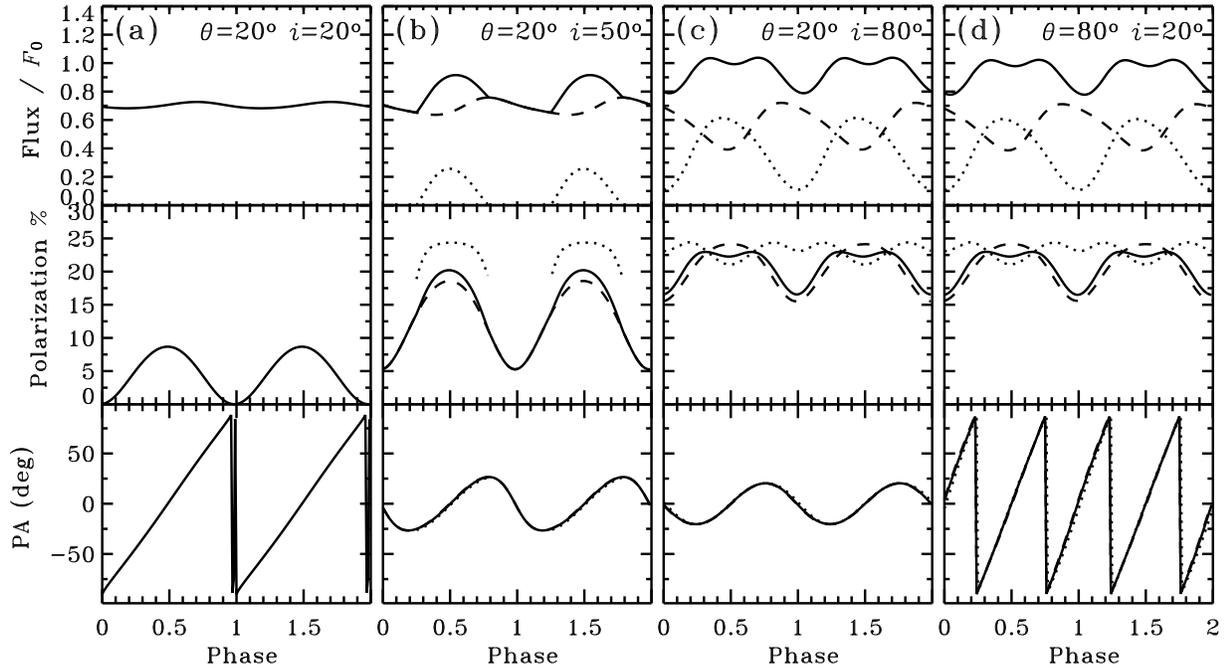,width=0.9\linewidth}}
\caption{Same as Fig.~\protect\ref{fig:nscat}, but
for scattering order $n=7$ and varying inclination $i$ and magnetic angle
$\theta$.
}
\label{fig:incl}
\end{center}
\end{figure*}

\begin{figure*}
\begin{center}
\centerline{\epsfig{file=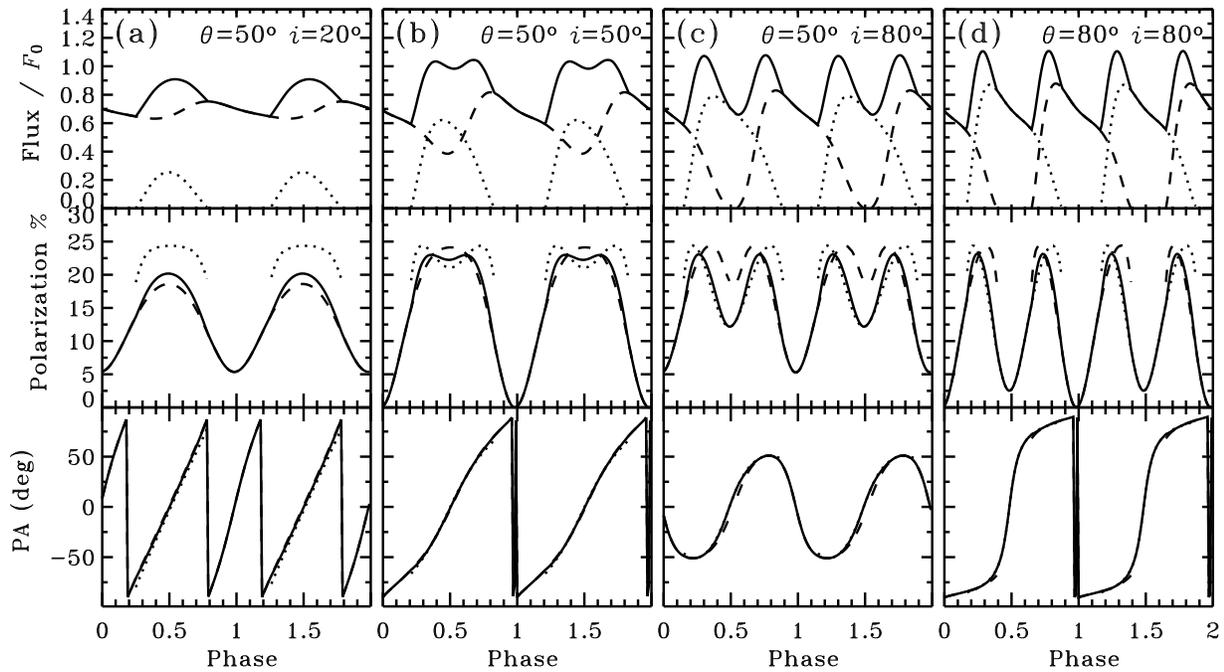,width=0.9\linewidth}}
\caption{Same as Fig.~\protect\ref{fig:incl},
for another set of parameters $\theta$ and  $i$.}
\label{fig:theta}
\end{center}
\end{figure*}

First,  in Fig.~\ref{fig:nscat} we show the dependence of the flux and
polarization profiles on the scattering order.
Radiation scattered once is beamed along the normal to
the slab similar to a black body radiation ($I(\mu)\approx \mbox{const}$).
The flux profiles are thus almost identical to those for the black body spots
(compare left panel in Fig.~\ref{fig:nscat} to the 2nd panel
of Fig.~\ref{fig:fluxbb}c) with the maximum close to the zero phase.
The polarization degree is always negative at first two scattering orders
 as can be seen from Fig.~\ref{fig:intpol}.
The PA is then varies around $90\degr$.
The polarization degree is larger for the secondary spot because we see it
at grazing angles.
However, the observed flux is larger for the primary spot
and hence the total degree of polarization is determined by the
primary. For each spot, the polarization is larger when the flux is lower
because polarization is a monotonic function of $\mu=\cos\alpha'$ which also
controls the flux.

\begin{figure*}
\begin{center}
\centerline{\epsfig{file=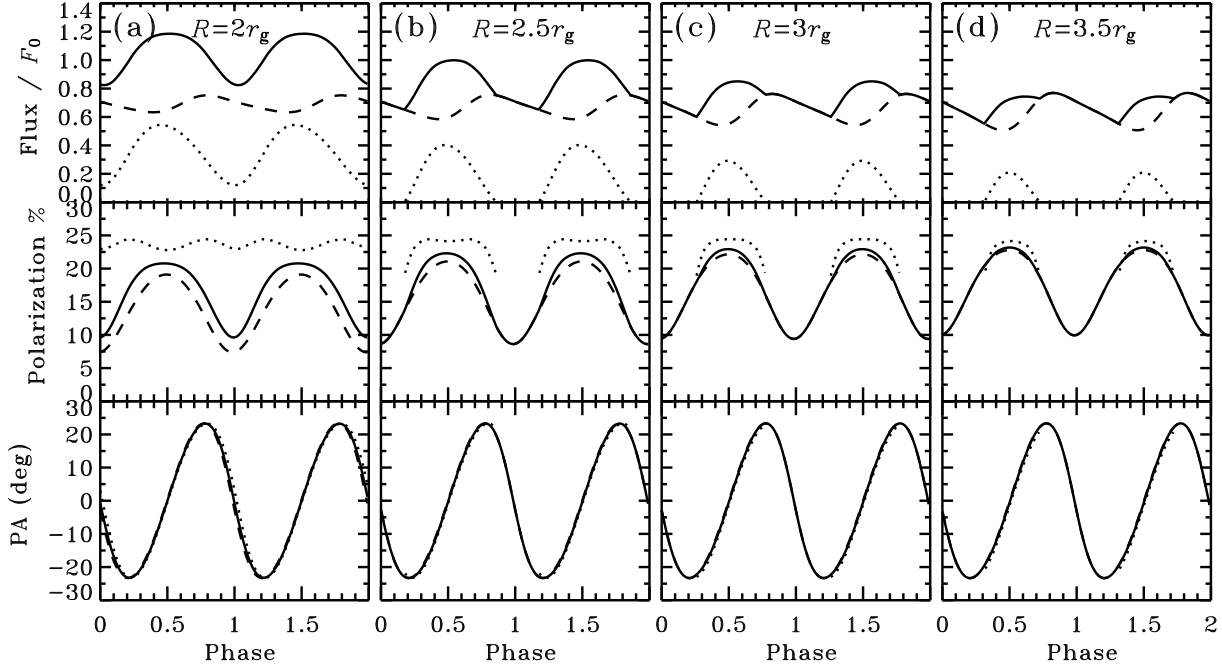,width=0.9\linewidth}}
\caption{Same as Fig.~\protect\ref{fig:incl}, but
for  varying neutron star compactnesses $R/\rg$.}
\label{fig:radrg}
\end{center}
\end{figure*}

With increasing $n$ the escaping flux becomes less dependent on the
polar angle resulting in the reduction of the variability for a primary spot.
The intensity  at large angles from the normal increases causing
a relative increase of the flux from the secondary spot
(which is visible at large angles). The total  flux is now peaked at
phase $0.5$ when the secondary has the maximum. At the third scattering,
the polarization degree defined by
Eq.~(\ref{eq:polslab}) changes sign for smaller polar angles, being still
negative for larger ones. The  PA for the
primary spot (which is observed closer to face-on) now varies around $0\degr$,
while the secondary spots, observed at grazing angles,
demonstrates a jump in the PA when polarization
changes the sign. For scattering order $n=7$
the polarization degree defined by Eq.~(\ref{eq:polslab})
is always positive, and thus, the PA for both spots varies around
zero degrees. The polarization degree is now not a monotonic function
of $\mu$, but reaches a maximum at $\mu\sim 0.2$ ($\alpha'\sim 80\degr$).
One sees that for the secondary spot $P$  does not
have the maximum when the flux is minimal since we see the spot at
grazing angles (see Fig.~\ref{fig:intpol}).
The hardly visible deviations in PA of the individual
spots are due to a combined effect of the time delay and a different
relativistic correction angle given by Eq.~(\ref{eq:pac}).

Let us represent the light curve for a given scattering order $n$
as a Fourier series
\be \label{eq:fourier}
F_n(\phi)=\overline{F_n}\ \left\{ 1+ \sum_{j=1} a_j(n)
\cos[ 2\pi j(\phi-\phi_j(n) ) ] \right\} ,
\ee
where $\overline{F_n}$ is the
mean flux. For a pure sine wave, the rms is $a_1/\sqrt{2}$. The
dependences of the rms and the variability amplitudes $a_1$, $a_2$
on the scattering order $n$ are shown in Fig.~\ref{fig:phaselag}
for three optical depths $\tau_0$. The amplitude of the
fundamental is generally a decreasing function of $n$ for one spot
(Fig.~\ref{fig:phaselag}a). If two antipodal spots are present
(Fig.~\ref{fig:phaselag}b), the rms has a minimum at $n=1-2$ when
$\tau_0=0.5-1$. At $n=0$, the black body is peaked as $I(\mu)={\rm
e}^{-\tau_0/\mu}$ along the normal to the slab producing large
variability. Radiation scattered once has now a  radiation pattern
which actually is close to a pure black body not covered by a hot
slab (see Fig.~\ref{fig:intpol} middle panel). This is known to
reduce the variability (B02, see Fig.~\ref{fig:nscat}a). At larger
$n$, radiation becomes beamed along the slab surface, secondary
spot becomes stronger  and variability increases
(Fig.~\ref{fig:nscat}c).

The phase lags $\phi_j(n)-\phi_j(0)$  relative to unscattered radiation,
i.e. $n=0$,
at the fundamental  frequency are shown in lower panels of Fig.~\ref{fig:phaselag}.
The negative lags mean that the harder photons are leading the softer
ones. With increasing scattering order the pulse
profile shifts to an earlier phase. For one spot (left panel), the
phase lag results from a combination of Doppler boosting and
a change in the emission pattern. The phase lag is then
a smooth function of $n$. It saturates after 5--10 scattering,
when the distribution of photons over the slab does not change anymore.
The lags are larger for smaller $\tau_0$ when radiation is stronger
beamed along the slab (fan-like emission pattern, see top panel in
Fig.~\ref{fig:intpol}).

For two spots (right panel),
there is a jump by almost half-a-period
in the lag dependence at small $n$ and $\tau_0$. This is again
a result of dramatic change in the radiation pattern at small $n$.
At $n=0$, the light curve is dominated by the primary spot and
the peak is close to $\phi=0$, while at larger $n$, the peak is
reached at phases where the secondary spot is visible.
For $\tau_0=3$, the radiation pattern changes gradually with $n$
and the phase lags also show a gradual evolution.

Figures \ref{fig:incl}--\ref{fig:theta} represent the light curves
and polarization profiles for $n=7$ and different
couples of $i$ and $\theta$  shown in Fig.~\ref{fig:class}.
We can see the dependence on the inclination
at the first three panels of both figures. When $i$ and $\theta$ are
small, we see only one spot (class I) and
the variability amplitude $\propto\sin i\ \sin\theta$
is also very small. Actually, it is much smaller than that for the
black body spot since the flux escaping from the slab after
a number of scatterings is more isotropic.
The polarization degree is small since
we look at the spot almost along the normal. If $i=\theta$,
then according to Eq.~(\ref{eq:papr}), the PA can be presented as
$\tan\chi \approx -1/[\cos i\ \tan (\phi/2)]$. For small $i$,
this reduces to a linear relation
$\chi\approx \phi /2 -\pi /2$ clearly visible in  Fig.~\ref{fig:incl}a,
while for larger $i$ the $\chi-\phi$ relation is more complicated
(see  Fig.~\ref{fig:theta}b,d).
For small $\beta$, the PA crosses the zero at phase
$\phi\approx -2\beta /(\sqrt{1-\rg/R} \sin i)$.

\begin{figure*}
\begin{center}
\centerline{\epsfig{file=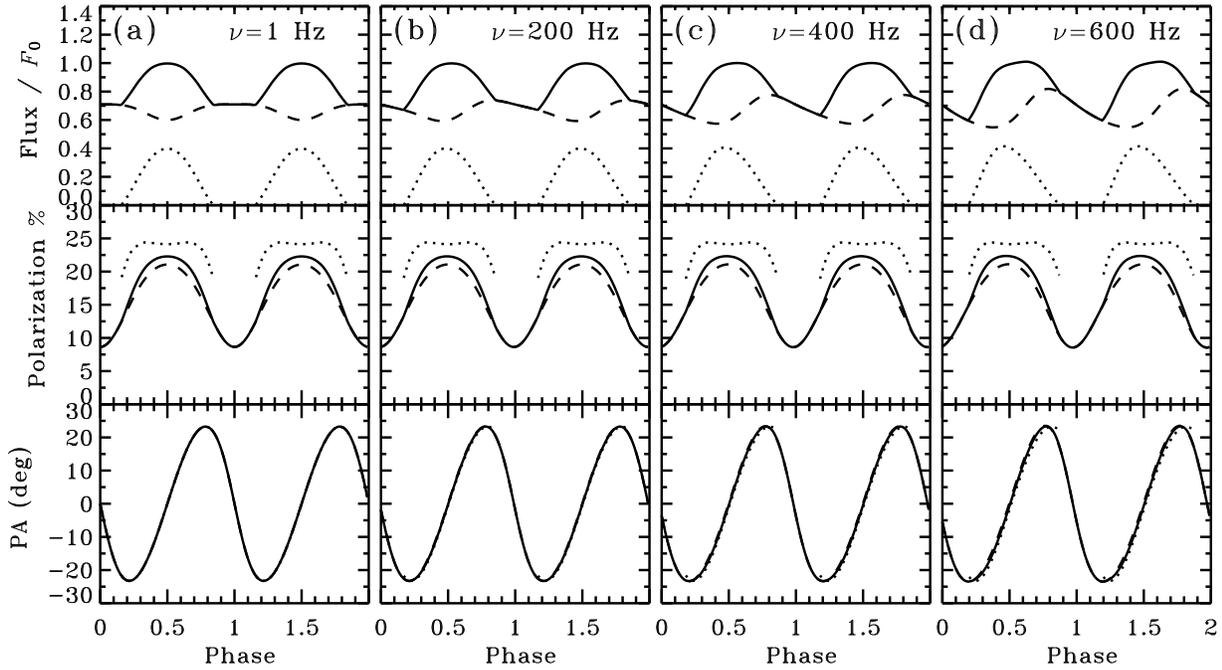,width=0.9\linewidth}}
\caption{Same as Fig.~\protect\ref{fig:incl}, but
for  varying  spin frequencies $\nu$.}
\label{fig:nu}
\end{center}
\end{figure*}

With increasing inclination, the second spot becomes visible
and the oscillation amplitude grows. In class IV,
it is actually larger than that for black body spots (compare
Fig.~\ref{fig:incl}c to the 3rd panel in Fig.~\ref{fig:fluxbb}a)
since the variations of the flux from the two spots do not exactly
cancel each other. The polarization also grows with $i$ since
we see the spots are larger angles. In Fig.~\ref{fig:incl}b
we also see that polarization from the primary spot
anti-correlates with corresponding flux, while for the secondary
spot they are correlated.
This is related to the fact that the primary spot is
observed at small polar angles $\mu\sim 1$ where flux and polarization
(see Fig.~\ref{fig:intpol}) are anti-correlated, while the secondary
spot is observed at larger angles where the relation reverses.
At larger inclinations (see Figs.~\ref{fig:incl}c, \ref{fig:theta}b,c)
the relation between
the polarization (from the secondary) and the flux is not monotonic since
we observe the spot at intermediate angles.
The PA varies around zero and is close to a sinusoid
$\chi\approx - (\tan\theta /\sin i) \sin\phi$ at large inclination and
small $\theta$ (see Figs.~\ref{fig:incl}b,c, \ref{fig:theta}c).
When both $i$ and $\theta$ are large (e.g. Fig.~\ref{fig:theta}d),
the flux and polarization degree vary mostly at the harmonic of the
spin frequency since the visibility conditions for both spots are
similar. The PA, however, varies with the spin frequency.

As we pointed out in \S~\ref{sec:bb}, the flux profile
is (almost) invariant to exchanging $i$ and $\theta$. The degree
of polarization is also invariant.
The PA, on the other hand, show completely different behavior:
when $i>\theta$ the variations are small,
but in the opposite case PA make a full turn of $360\degr$
(compare Fig.~\ref{fig:incl}b to Fig.~\ref{fig:theta}a and
Fig.~\ref{fig:incl}c to Fig.~\ref{fig:incl}d).
At $i<\theta$, the PA is almost a linear function of the phase,
with the derivatives
\be
\left. \frac{\d \chi}{\d\phi} \right|_{\phi=0} =
\frac{\sin\theta}{\sin(\theta-i)}, \quad
\left. \frac{\d \chi}{\d\phi} \right|_{\phi=\pi} =
\frac{\sin\theta}{\sin(\theta+i)}.
\ee

\begin{figure*}
\begin{center}
\centerline{\epsfig{file=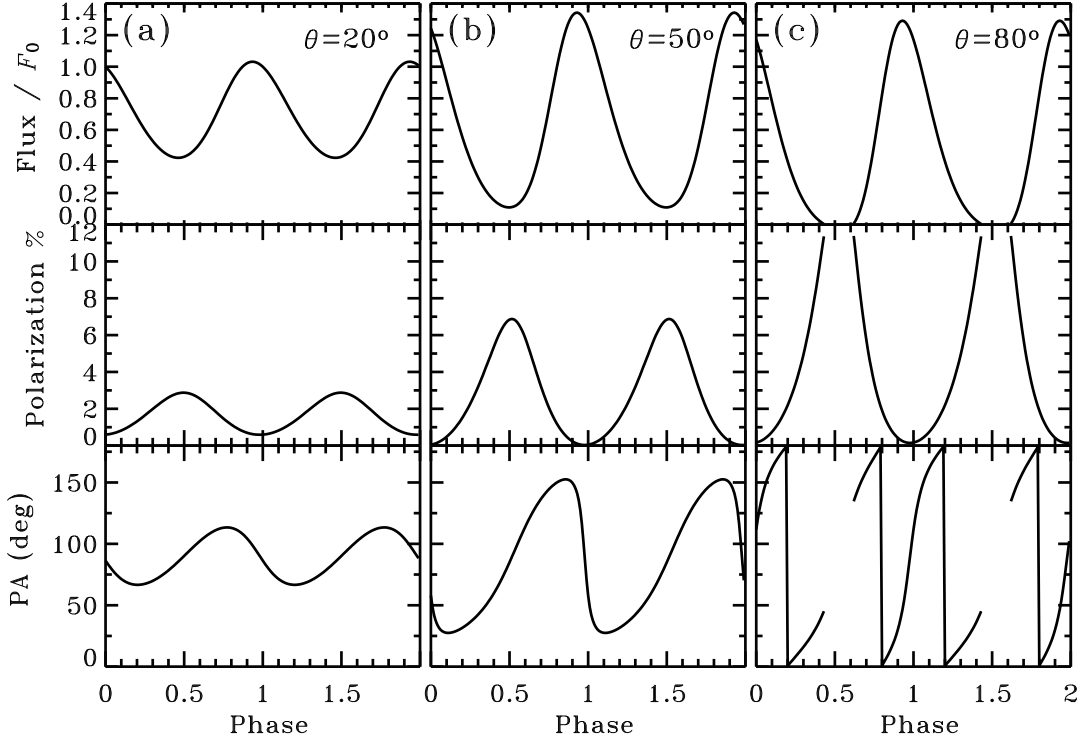,width=0.8\linewidth}}
\caption{Light curves, polarization degree and polarization angle
expected from an X-ray burst. Semi-infinite electron scattering
atmosphere is considered. The parameters are $M=1.4\msun$, $i=60\degr$,
$R=2.5\rg$, $\nu=400$ Hz. }
\label{fig:burst}
\end{center}
\end{figure*}

The compactness of the neutron star influences the gravitational
bending and changes the angle the spots are observed at.
For small radius, bending is strong and both spots can be visible all
the time (class IV light curve, see Fig.~\ref{fig:radrg}a), while at
larger radius
the light curve class can change to II or even I (Fig.~\ref{fig:radrg}b-d).
The polarization degree slightly increases with $R$ (effect is
clearly visible for the primary spot), since for the given angle $\psi$,
$\cos\alpha\approx \rg/R + (1-\rg/R)\cos\psi$
is a monotonically decreasing function of $R$.
The effect of the radius on the PA is tiny since only
(small) correction angle (\ref{eq:pac}) depends on it.

The influence of the neutron star spin is shown in Fig.~\ref{fig:nu}.
Increasing $\nu$ modifies mostly the flux profile, since it is
proportional to the Doppler factor in power $\Gamma+3$.
The peak shifts towards the phase where the Doppler factor reaches
the maximum (a quarter of the period before the peak of the
projected area). The pulse profiles thus becomes more skewed.
The polarization degree depends on $\cos\alpha'=\Dop\cos\alpha$,
which varies only slightly with the spin. The PA is affected
through the correction angle which is a linear function
of $\nu$ (see Eq.~(\ref{eq:pac})) and is small.
The time delay slightly shifts all the profiles, especially
those for the secondary spot.

\subsection{Phase lags in accreting millisecond pulsars}

The detailed fitting of the light curves predicted by our
models to the data is outside the scope of this paper
(see e.g. PG03 for applications to SAX~J1808.4$-$3658).
We just compare here the general evolution of the pulse
profile with energy which is quantified by the phase lags.
We show the phase lags observed from the first accreting ms pulsar
SAX~J1808.4$-$3658 (Cui et al. \cite{cui98}; Gierli\'nski et al. \cite{gier02})
in the lower panel of Fig.~\ref{fig:phaselag}a.
One sees that the observed behavior is consistent with
the model predictions for  a single spot and
optical depth of the slab of $\tau_0=1$.
Such an optical depth is in turn consistent with the observed X-ray spectra from
the source (PG03).
SAX~J1808.4$-$3658 shows almost sinusoidal variations which cannot
be described by two-spot model. As is shown in PG03, most probably,
the second antipodal spot is blocked from the observer by the accretion disk
which inner radius has to be smaller than a few neutron star radii.

The second accreting ms pulsar XTE~J1751$-$305
has pulse profiles which are almost identical to those of
SAX~J1808.4$-$3658 with somewhat smaller rms amplitude (Markwardt et al. \cite{mark02}).
Again, only one spot seems to be visible.
The spectral analysis (Gierli\'nski  \& Poutanen, in preparation)
shows that the optical depth of the Comptonizing medium
is $\tau_0\sim2$.  According to the model,
the phase lags then are smaller than those for
$\tau_0\sim 1$. Observations confirm (diamonds at the lower panel
Fig.~\ref{fig:phaselag}a) that this indeed is the case.
One should note here that XTE~J1751$-$305 has a slightly different pulsar
period (2.3 ms vs 2.5 ms in SAX~J1808.4$-$3658), therefore the
model predictions should be changed accordingly.

We emphasize that the model phase lags are computed relative
to the zeroth scattering order (black body), while the observed
lags are computed relative to the $\sim 3$ keV energy band where
photons scattered in the hot slab can
provide significant contribution. It is encouraging that
this simplified model provides a good description of  the data.
At this this stage, however, a detailed  comparison seems unjustified.

\subsection{X-ray bursts}

The  X-ray bursts often show coherent oscillations (see
Strohmayer \& Bildsten \cite{sb03}).
The energy dissipation takes place deep in the
electron scattering atmosphere, and thus we can assume that
Thomson optical depth is infinite.
The escaping intensity in that case is $I(\mu)\propto1+2\mu$ and
polarization is 11.7\% at maximum when the spot is viewed  edge-on
(see  dashed curves in Fig.~\ref{fig:intpol}).
The predicted pulse profiles and behavior of polarization degree
and angle are shown in Fig.~\ref{fig:burst}.
Since the radiation pattern is peaked sharper along the normal
than that of the black body,
the variability amplitude is larger (compare Fig.~\ref{fig:burst}a
to 2nd panels in Fig.~\ref{fig:fluxbb}a,b). The polarization
is increasing with the spot colatitude $\theta$ reaching
the maximum of $\sim 12\%$ close to the eclipses.
The PA varies around $90\degr$ as the electric vector
is predominantly perpendicular to the meridional plane
and the amplitude grows with $\theta$.

\section{Summary}

In this work we have developed a formalism to compute
waveforms and polarization profiles from
accreting millisecond pulsars and coherent oscillations in X-ray bursts.
The radiation from emission region was assumed to be produced
as a results of Compton scattering in the accretion shock
of Thomson optical thickness $\sim1$
(in the case of accretion-powered sources) or optically thick
electron atmosphere (in the case of X-ray bursts).
The observed flux and polarization are affected by special
and general relativistic effects, such as Doppler boosting,
relativistic rotation of the polarization plane, light bending, and
gravitational redshift.
The developed technique is accurate and extremely efficient from
the computational point of view.

Future polarimetric observations in the X-ray domain will serve
as a powerful tool in determining the geometry of the emission region
in rapidly rotating neutron stars showing coherent millisecond
oscillations. Our simulations are the first step in theoretical
modelling of these sources.

\begin{acknowledgements}
This work was supported by the Academy of Finland,
the Jenny and Antti Wihuri Foundation, and the NORDITA Nordic
project on High Energy Astrophysics.

\end{acknowledgements}

\end{document}